\documentclass[preprint2,longabstract]{aastex}

\usepackage{graphicx}
\shorttitle{APASS Landolt-Sloan BVgri photometry of RAVE stars}
\shortauthors{Munari et al.}

\begin{document}

\title{APASS Landolt-Sloan BVgri photometry of RAVE stars. \\I. Data, effective temperatures and reddenings}

\author{U.~Munari}
\affil{INAF Osservatorio Astronomico di Padova, 36012 Asiago (VI), Italy}

\author{A.~Henden} 
\affil{AAVSO, Cambridge, Massachusetts, USA}

\author{A.~Frigo}
\affil{ANS Collaboration, c/o Astronomical Observatory, Padova, Italy}

\author{T.~Zwitter}
\affil{Faculty of Mathematics and Physics, University of Ljubljana, 1000 Ljubljana, Slovenia}

\author{O.~Bienaym\'e}
\affil{Observatoire Astronomique, Universit\'e de Strasbourg, CNRS, 11 rue de l'universit\'e 67000 Strasbourg, France}

\author{J.~Bland-Hawthorn}     
\affil{Sydney Institute for Astronomy, University of Sydney, NSW 2006, Australia}

\author{C.~Boeche}
\affil{Astronomisches Rechen-Institut, Zentrum f\"ur Astronomie, M\"onchhofstr. 12-14, 69120 Heidelberg}

\author{K.C.~Freeman}		 
\affil{Mount Stromlo Observatory, RSAA, Australian National University, Weston Creek, Canberra, ACT 2611, Australia}

\author{B.K.~Gibson}
\affil{Jeremiah Horrocks Institute, University of Central Lancashire, Preston, PR1 2HE, UK}
\affil{Inst. for Computational Astrophysics, Saint Mary’s University, Halifax, NS, BH3 3C3, Canada}

\author{G. Gilmore}
\affil{Institute of Astronomy, Cambridge University, Madingley Road, Cambridge CB3 0HA, UK}

\author{E.K.~Grebel}		
\affil{Astronomisches Rechen-Institut, Zentrum f\"ur Astronomie, M\"onchhofstr. 12-14, 69120 Heidelberg}

\author{A.~Helmi}	       
\affil{Kapteyn Astronomical Institute, University of Groningen, PO Box 800, 9700 AV Groningen, the Netherlands}

\author{G. Kordopatis}
\affil{Institute of Astronomy, Cambridge University, Madingley Road, Cambridge CB3 0HA, UK}

\author{S.E.~Levine} 
\affil{Lowell Observatory, Flagstaff, Arizona, USA}

\author{J.F.~Navarro}  	 
\affil{Department of Physics and Astronomy, University of Victoria, Victoria, BC, Canada V8P 5C2}

\author{Q.A.~Parker}  
\affil{Department of Physics \& Astronomy, Macquarie University, NSW 2109, Australia}
\affil{Australian Astronomical Observatory, PO Box 915, North Ryde, NSW 1670, Australia}

\author{W.~Reid}
\affil{Research Centre in Astronomy, Astrophysics and Astrophotonics, Macquarie University, Australia}
\affil{Department of Physics \& Astronomy, Macquarie University, NSW 2109, Australia}

\author{G.M.~Seabroke} 	 
\affil{Mullard Space Science Laboratory, University College London, Holmbury St Mary, Dorking, RH5 6NT, UK}

\author{A.~Siebert}
\affil{Observatoire Astronomique, Universit\'e de Strasbourg, CNRS, 11 rue de l'universit\'e 67000 Strasbourg, France}

\author{A.~Siviero}
\affil{INAF Osservatorio Astronomico di Padova, 36012 Asiago (VI), Italy}

\author{T.C.~Smith}
\affil{Dark Ridge Observatory, Weed, New Mexico, USA}

\author{M.~Steinmetz}
\affil{Leibniz-Institut für Astrophysik Potsdam (AIP), An der Sternwarte 16, D-14482 Potsdam, Germany}

\author{M.~Templeton}
\affil{AAVSO, Cambridge, Massachusetts, USA}

\author{D.~Terrell}
\affil{Southwest Research Institute, Boulder, Colorado, USA}

\author{D.L.~Welch}
\affil{McMaster University, Hamilton, Ontario, Canada}

\author{M.~Williams}
\affil{Leibniz-Institut für Astrophysik Potsdam (AIP), An der Sternwarte 16, D-14482 Potsdam, Germany}
\and
\author{R.F.G.~Wyse}
\affil{Johns Hopkins University, Homewood Campus, 3400 N Charles Street, Baltimore, MD 21218, USA}



\begin{abstract}
We provide APASS photometry in the Landolt $B$$V$ and Sloan
$g'$$r'$$i'$ bands for all the 425,743 stars included in the
latest 4th RAVE Data Release.  The internal accuracy of the APASS
photometry of RAVE stars, expressed as error of the mean of
data obtained and separately calibrated over a median of 4
distinct observing epochs and distributed between 2009 and
2013, is 0.013, 0.012, 0.012, 0.014 and 0.021 mag for
$B$,$V$,$g'$,$r'$ and $i'$ band, respectively.  The equally
high external accuracy of APASS photometry has been verified
on secondary Landolt and Sloan photometric standard stars not
involved in the APASS calibration process, and on a large
body of literature data on field and cluster stars,
confirming the absence of offsets and trends.  Compared with
the Carlsberg Meridian Catalog (CMC-15), APASS astrometry of
RAVE stars is accurate to a median value of 0.098 arcsec. 
Brightness distribution functions for the RAVE stars have
been derived in all bands.  APASS photometry of RAVE stars,
augmented by 2MASS $J$$H$$K$ infrared data, has been $\chi^2$
fitted to a densely populated synthetic photometric library
designed to widely explore in temperature, surface gravity,
metallicity and reddening. Resulting $T_{\rm eff}$ and
$E_{B-V}$, computed over a range of options, are provided and
discussed, and will be kept updated in response to future
APASS and RAVE data releases. In the process it is found that
the reddening caused by an homogeneous slab of dust,
extending for 140 pc on either side of the Galactic plane and
responsible for $E^{poles}_{B-V}$=0.036$\pm$0.002 at the
galactic poles, is a suitable approximation of the actual
reddening encountered at Galactic latitudes
$|b|$$\geq$25$^\circ$.
\end{abstract}



\section{Introduction}

  \begin{figure*}[!Ht]
     \centering
     \includegraphics[angle=270,width=16cm]{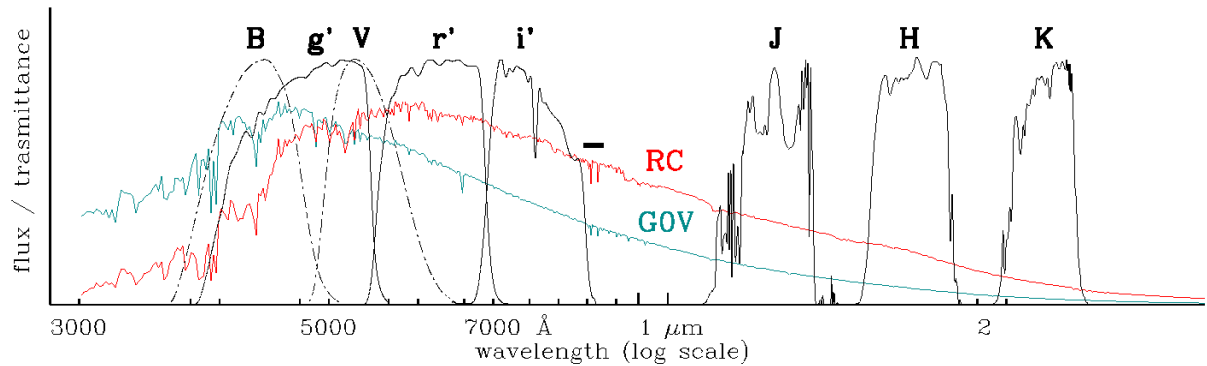}
     \caption{The spectral energy distributions of a G0 main sequence and of
     a Red Clump star, over-plotted with the transmission profiles of Landolt
     $BV$, Sloan $g'r'i'$ and 2MASS $JHK$ bands
     (including telluric absorptions).  The thick horizontal dash
     at 8600~\AA\ marks the wavelength range covered by RAVE spectra.
     \label{fig1}}
  \end{figure*}

RAVE (RAdial Velocity Experiment) is a digital spectroscopic survey of stars
in the nominal magnitude interval $9\leq I_{\rm C} \leq12$, distributed over
the whole southern sky.  Spectra have been recorded over the 8400-8800~\AA\
range, at an average resolving power $\sim$6500, with the AAO (Australian
Astronomical Observatory) UK~Schmidt telescope feeding light to a
spectrograph via the 6-degree Field (6dF) 150 fiber positioner.  The primary
science driver of the survey is the investigation of the structure and
evolution of the Milky Way through the determination of radial velocities,
chemistries, temperatures, and gravities for a large number of stars.  The
survey data releases \citep{ste06,zwi08,sie11} have followed an incremental
scheme, with the last Data Release N.4 \citep[DR4][]{kor13} covering
482\,194 spectra for 425\,743 individual stars.  RAVE data have so far
directly supported about fifty refereed papers.

Accurate photometry, based on bands spanning the optical and near infrared
wavelength ranges, is an essential support to spectroscopic surveys like
RAVE.  Such photometry provides, among other things: ($i$) the brightness
distribution functions of the survey targets, ($ii$) the effective
temperature, useful {\em per se} or as a prior to the spectroscopic
determination of atmospheric parameters, ($iii$) the reddening, ($iv$) the
observed brightness against which to derive the distance modulus ($m-M$)
to the survey targets, ($v$) the color discriminant to segregate single
normal stars from binaries and peculiar stars, and ($vi$) when multi-epoch
photometry is available, information on the target intrinsic variability. 
So far, only limited photometric information in the optical has been
available for RAVE stars, as reviewed below. The goal of this paper is
to provide accurate multi-band optical photometry for all RAVE stars and
to combine it with near-IR 2MASS photometry.

A total of 251\,626 RAVE DR4 stars have a match in the Tycho-2 $B_T$,$V_T$
catalog.  Unfortunately, the Tycho-2 photometric errors rapidly
degrade toward the faint survey limit, and over the range of brightness of
the bulk of RAVE stars, Tycho-2 data (especially the $B_T$ band) are
virtually useless.  The median value of quoted Tycho-2 errors for RAVE
stars is $\sigma (B_T)= 0.15$ and $\sigma (V_T)=0.09$~mag, for a combined
$\sigma (B-V)_T\sim 0.18$ which is far larger than the reddening affecting
the majority of RAVE stars, or larger than the color spread corresponding to
the combined uncertainty in the RAVE atmospheric parameters ($T_{\rm eff}$,
$\log g$, [M/H]).

A greater number of RAVE DR4 stars, 407\,061 in total, have a counterpart in
the DENIS $I$,$J$,$K$ catalog.  DENIS $J$ and $K$ photometry is of limited
precision, the median value on RAVE stars being $\sigma (J)$=0.07 and
$\sigma (K)$=0.07 mag (three times worst that corresponding 2MASS values),
with $\sigma (J-K)=0.10$ on the combined color index ($\sigma (J-K)=0.033$
for 2MASS).  Such an uncertainty on the color index corresponds to an
uncertainty of 15\% in the effective temperature of a red giant star or 0.20
mag in $E_{B-V}$, spoiling DENIS $J$,$K$ photometry of any practical value
in supporting RAVE survey.  DENIS $I$ photometry, even if reported to a
greater precision (a median of $\sigma (I)=0.03$~mag on RAVE stars), is
affected by saturation problems for bright RAVE targets, setting in at
undocumented different levels of brightness depending on the location on the
sky \citep[cf.][]{kor13}.  Furthermore, the transmission profile of
DENIS $I$ band is not accurately known, and thus the observed complex offset
from Landolt's and Cousins' standard stars cannot be safely compensated for
during modeling and analysis of the data.

  \begin{figure}[!Ht]
     \centering
     \includegraphics[width=7.5cm]{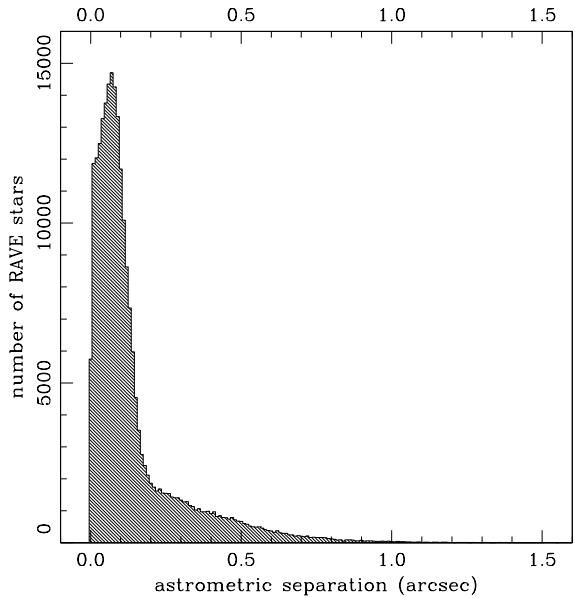}
     \caption{Distribution of the differences (in arcsec) between the
     astrometric positions derived by APASS for the RAVE stars and the
     positions listed for the same stars in the latest issue (CMC-15 of
     2013) of the Carlsberg Meridian Catalog.\label{fig2}}
  \end{figure}

A further source of photometry for a consistent fraction of RAVE stars is
the $r'$ band data obtained as a by-product during the astrometric operation
of the Carlsberg Meridian Circle in La Palma, covering the sky between
declinations $-40 \leq \delta \leq +50$. The final Carlsberg Meridian
Circle data release, CMC-15 of 2013, contains counterparts to 228\,805 RAVE DR4
stars. The original CMC observations did not include a second photometric
band, necessary for the transformation from the local to the standard
system, which was achieved by importing in the process the bluer color
($B_T - V_T$) from Tycho-2 catalog. The median value of the CMC-15
reported internal error on $r'$ band data for RAVE DR4 stars is 0.026 mag.
The CMC-15 $r'$ band data will be further considered in sect. 2.5.

On the near-IR side, 2MASS $J$, $H$, and $K$ magnitudes are available for
all RAVE DR4 stars.  The median value of the quoted errors for RAVE stars is
$\sigma (J)$=0.024, $\sigma (H)$=0.025, $\sigma (K)$=0.023 mag, much better
than for the corresponding DENIS bands.  However, even if 2MASS $J$,$H$,$K$
data are available for all RAVE stars, and they are of reasonable accuracy,
nevertheless they map the Rayleigh-Jeans tail of the energy distributions of
typical RAVE stars (cf.  Figure~1), which provides only a limited leverage
in term of diagnostic capability.  In fact, the 10$-$90 percentile interval
in the color distribution of RAVE stars corresponds to just
$\Delta (J-K)=0.87$~mag.

The aim of the present paper is to provide accurate photometry for all RAVE
DR4 stars in the Landolt $B$,$V$ and Sloan $g'$,$r'$,$i'$ photometric bands,
as obtained during the APASS survey 
\citep[Aavso Photometric All-Sky Survey;][]{hen14,hen14a}. Combining with 2MASS
$J$,$H$,$K$ data in the infrared, APASS photometric bands encompass the
majority of the energy distribution of typical RAVE stars, as shown
schematically in Figure~1.  The plan of this paper is first to present the
APASS data on RAVE stars and test their photometric and astrometric quality. 
Photometric degeneracy between temperature and reddening is then discussed,
and a simple model for galactic extinction is introduced and its parameters
calibrated against APASS+2MASS data for standard stars.  Photometric
temperatures are derived, first for a set of standard stars to assess their
accuracy, and then for the whole body of RAVE stars.  These photometric
temperatures are compared with those derived spectroscopically, and the
result discussed.  Finally, the $E_{B-V}$ reddening is derived for all RAVE
DR4 stars both from unconstrained fit as well as fixing the atmospheric
parameters to the values derived by RAVE spectra.

Among the goals of future papers in this series, there will be the
derivation of a Galactic 3D map for the interstellar $E_{B-V}$ reddening
here derived for RAVE DR4 stars, the photometric characterization of RAVE
DR4 peculiar stars, the study of the photometric variability of RAVE DR4
peculiar stars by accessing the individual epoch photometric data, the
use of supplementary observations in the $u'$, $z'$ and $Y$ filters, and
extension to brighter magnitudes than the current APASS saturation limits.

\section{The data}

\subsection{The APASS survey}

  \begin{figure*}[!Ht]
     \centering
     \includegraphics[height=16.0cm,angle=270]{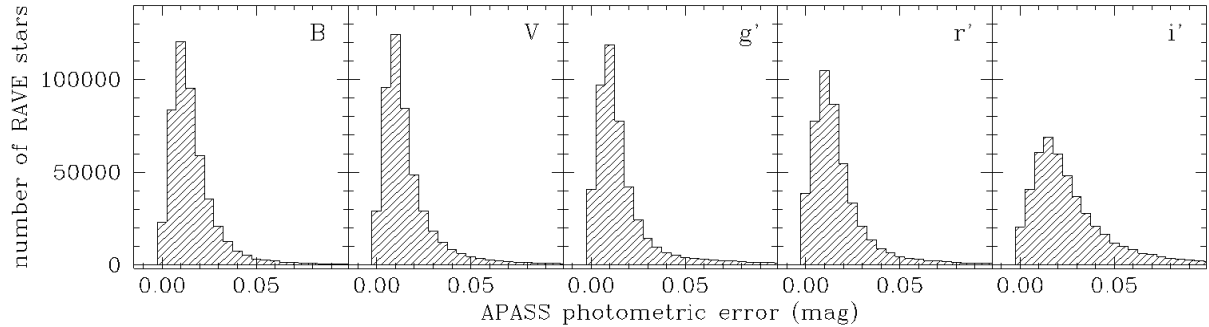}
     \caption{Distribution of errors of APASS $B V g' r' i'$ magnitudes for 
     RAVE stars (errors of the mean).\label{fig3}}
  \end{figure*}

  \begin{figure}[!Ht]
     \centering
     \includegraphics[width=7.5cm]{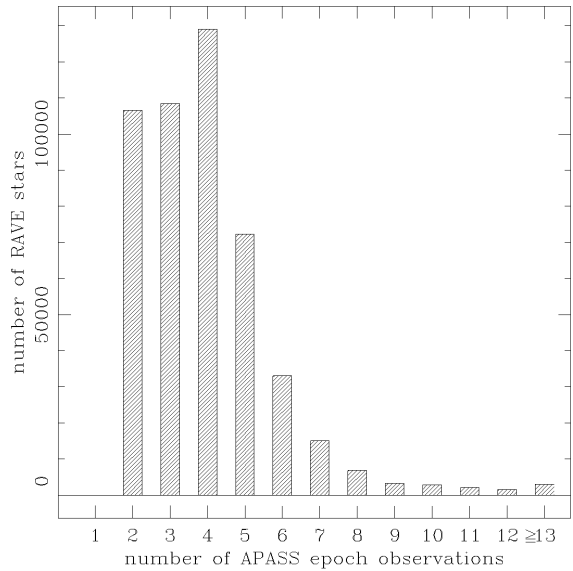}
     \caption{Number of independent epochs in APASS observations of RAVE
     stars.\label{fig4}}
  \end{figure}

The $B V g' r' i'$ photometric data of RAVE stars discussed in this
paper have been obtained as part of the ongoing APASS survey 
\citep{hen14,hen14a}.  The all-sky APASS survey is carried out
to fill the gap existing between the completeness limit of the all-sky
Tycho-2 $B_T V_T$ survey ($V_T \sim 11$) and the saturation limit
($V \sim 15$) of planned or already operating near-IR and optical
photometric surveys like SDSS, PanSTARRS, LSST, SkyMapper, VISTA, VST, etc. 
APASS results are already used as the photometric source catalog for the US
Naval Observatory CCD Astrograph Catalog (UCAC-4), and for the Harvard College
DASCH program (which is scanning the Harvard plate stacks and measuring the
magnitude of all detected stars).

The APASS photometric survey covers the whole sky, from Pole to Pole, with
ongoing observations from CTIO (Chile), for the southern hemisphere, and New
Mexico for the northern counterpart.  At both sites, a pair of twin remotely
controlled, small telescopes obtain simultaneous CCD observations during
dark- and grey-Moon time over five optical bands: $B$, $V$ 
\citep[tied to the equatorial standards of][]{lan09} and $g'$,$r'$,$i'$ bands
(tied to the 158 primary standards given by \citet{smi02}, that define
the Sloan photometric system).  The telescopes are 20-cm f/3.6 astrographs
feeding Apogee U16m cameras (4$096 \times 4096$ array, 9~$\mu$m pixels), that
cover a field 2.9~deg wide with a 2.6~arcsec/pix plate factor.  The
photometric filters are of the dielectric multi-layer type and are produced
by Astrodon.  Transmission curves and photometric performances of Astrodon
filters are discussed and compared to more conventional types of photometric
filters in \citet{mun12} and \citet{mun12a}.  On average
80 fields are observed per night at each APASS location, 20 of them being
standard fields (Landolt, Sloan) well distributed in time and airmass.

The APASS observations are obtained with fixed exposure times (different and
optimized for each photometric band), set to detect $V$=17 stars at S/N=5 on
a single exposure.  Stars brighter than $V$=10 may saturate under optimal
seeing conditions.  APASS has also recently begun using the bright-Moon
periods to obtain shorter exposure observations in order to push the
saturation limit upward to 7.5 mag stars, and to observe also in the $Y$
band at 1.035 $\mu$m \citep[defined by][]{hil02,sha10} and Sloan $u'$, $z'$ 
bands.

\subsection{Astrometric matching}

The astrometric matching to the APASS catalog has been performed by taking
position and proper motions of RAVE stars as listed in DR4 and projecting
them to the 2013 APASS epoch.  The matching radius has been set to 3.0
arcsec.  All 425\,743 RAVE DR4 stars turned out to have their APASS
counterpart.

As a test on the accuracy of APASS astrometry (which is calibrated
against UCAC-4), we have searched the final CMC-15 release (2013) of the
Carlsberg Meridian Catalog for the positions of RAVE stars and computed the
separation between the APASS and Carlsberg positions (equinox J2000,
individual epochs varying from 1999 to 2005).  Their distribution is
presented in Figure~2.  It has its peak at a separation of 0.072 arcsec and
the median at 0.098 arcsec, confirming the high astrometric accuracy of
APASS products.

\subsection{Internal photometric accuracy}

Each APASS observing night is independently calibrated in all-sky mode
against the Landolt and Sloan standards observed that same night over a
large range of airmass.  The photometric quality of a given night can be
evaluated only {\em a posteriori} during the data reduction process, and
only data obtained during photometric nights are used to build the APASS
catalog.  The difference in magnitude for data on the same stars obtained in
separate nights (usually months apart), is therefore a reliable estimate of
the {\it total} internal errors affecting APASS photometry.  Their
distribution for RAVE stars is presented in Figure~3, and the distribution
of RAVE stars as function of the number of epoch APASS observations is given
in Figure~4 (on average, a RAVE star has been observed on 4 independent
epochs by APASS, and none at less than 2 epochs).  The errors corresponding
to the tails of the distributions in Figure~3 are actually an over-estimate
of the true internal photometric errors.  In fact, in building Figure~3 all
stars are treated as non-variable, but as shown by the Hipparcos
satellite about 10\% of all stars it observed did show a variability of at
least 0.01 mag amplitude.  Median values of the error distributions in
Fig.~3, and the corresponding 2MASS values are:
\begin{eqnarray}
err(B)  &=& 0.013 {\rm ~~mag} \nonumber \\
err(V)  &=& 0.012 {\rm ~~mag} \nonumber \\
err(g') &=& 0.012 {\rm ~~mag} \nonumber \\
err(r') &=& 0.014 {\rm ~~mag} \\
err(i') &=& 0.021 {\rm ~~mag} \nonumber \\
err(J)  &=& 0.024 {\rm ~~mag} \nonumber \\
err(H)  &=& 0.025 {\rm ~~mag} \nonumber \\
err(K)  &=& 0.023 {\rm ~~mag} \nonumber
\end{eqnarray}
where the larger value for the $i'$ band (compared to $B$,$V$,$g'$ and $r'$
bands) is mainly due to the variable intensity of the strong telluric
absorptions present over its pass-band.  By the time the final products of APASS
survey will be released and the number of observing epochs doubled, the
above errors are expected to reduce by $\sim 30$\%.

  \begin{figure}
     \centering
     \includegraphics[width=7.5cm]{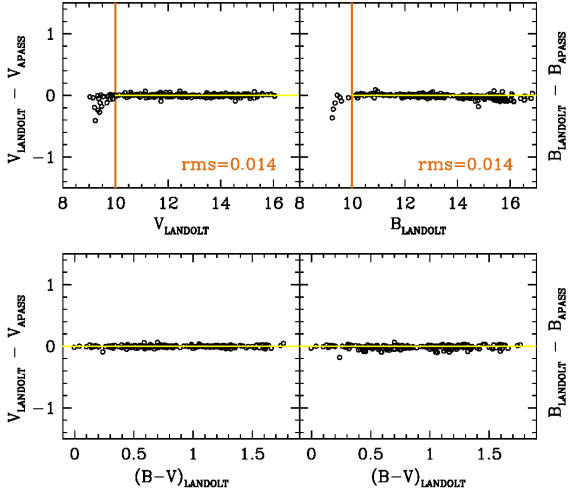}
     \caption{Comparison between the source $B$,$V$ data of the \citet{lan09} 
     equatorial standards and the corresponding APASS values. 
     The vertical line marks where saturation of brighter stars sets in
     (because of APASS fixed exposure times) under excellent seeing
     conditions. Saturated stars from top panels are not plotted on lower
     panels.\label{fig5}}
  \end{figure}

  \begin{figure}
     \centering
     \includegraphics[width=7.5cm]{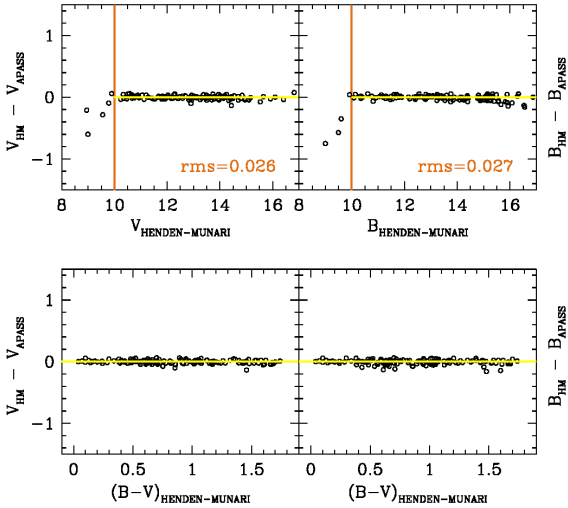}
     \caption{Comparison between the $B$,$V$ data of 323 secondary, low
     Galactic latitude Landolt standards calibrated by \citet{hen00,hen01,hen06} 
     and the corresponding APASS values. Saturated stars 
     from top panels are not plotted on lower panels.\label{fig6}}
  \end{figure}

\subsection{Saturation limits}

A necessary test is to evaluate at what magnitude saturation sets in
(because of the fixed exposure time over the whole sky), and therefore to
place a safe upper brightness limit to the use of APASS data to model
properties of RAVE stars.  The test is carried out in Figure~5, where the
tabular value of 245 Landolt primary standards is compared to that derived
by APASS (which obviously uses only non-saturated Landolt standards to
perform its calibration).  Note that APASS uses the same 14-arcsec diameter
digital aperture that matched the Landolt 14-arcsec photoelectric photometer
aperture, so any neighboring star included in Landolt's aperture is also
included in the APASS photometry.  Figure~5 shows that most of the $B$,$V$
observations are unsaturated up to 9.5 mag, but excellent seeing can cause
this limit to shift down to 10.0 mag.  A similar exercise for Sloan 158
primary standards shows that $i'$ band is unsaturated up to 9.0 mag, the
$r'$ band up to 9.5 mag, while the limit for the $g'$ band is 10.0 mag. 
These results perfectly match those independently derived, from a different
and independent set of APASS data, during the search for and the
characterization of RR Lyr variables potentially associated with the
Aquarius stream \citep{mun14}, a Galactic stream originating from a
disrupting globular cluster that was first identified on RAVE data 
\citep{wil11,wyl12}.

To have a safe margin in all bands and for all stars, we will limit the
analysis in the rest of this paper to stars with an APASS magnitude
$\geq 10.0$ in the $B$,$V$,$g'$,$r'$ bands and $\geq 9.5$ in $i'$ (only 1.8\%
of the total number of RAVE DR4 stars is consequently ignored in the
following).  As mentioned in the introduction, APASS is currently using the
bright Moon nights to perform a shallower parallel survey aiming to push the
saturation limit up to 7.5 mag stars.  The APASS data for the brighter RAVE
stars will be investigated when ready, thus later in this series of papers.

  \begin{figure}[!Ht]
     \centering
     \includegraphics[width=7.5cm]{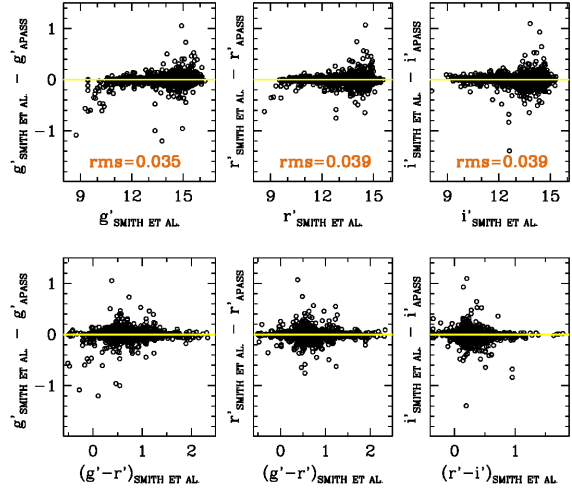}
     \caption{Comparison between the $g' r' i'$ data of 2988 
     secondary Sloan standards provided by \citet{smi02}
     and the corresponding APASS values.\label{fig7}}
  \end{figure}

  \begin{figure}[!Ht]
     \centering
     \includegraphics[width=7.5cm]{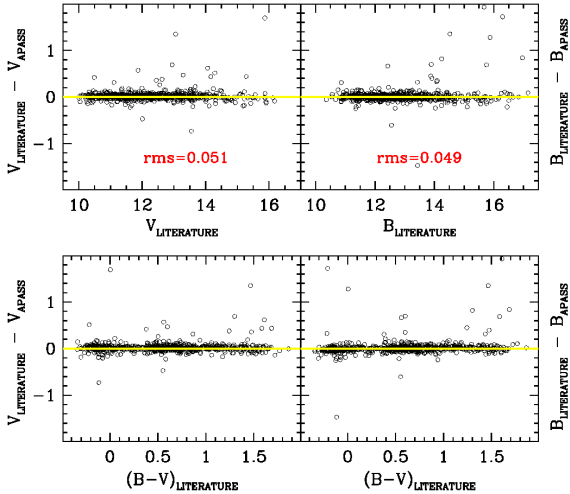}
     \caption{Comparison between literature $B$,$V$ data on field
     stars compiled by \citet{mer94} and the corresponding
     APASS values.\label{fig8}}
  \end{figure}

\subsection{External photometric accuracy}

Even if the small internal errors already make APASS the most accurate
all-sky, multi-band photometric survey in the optical wavelength range, what
really matters are the external errors.  To estimate them, we compare APASS
data to available literature. We begin with photometric standard stars not
used in the calibration of APASS data (to avoid circular arguments), and then
move to consider field and cluster stars.

\citet{hen00,hen01,hen06} have calibrated a series of secondary
Landolt standards in 81 low galactic latitude fields.  These secondary
standards are {\it not} used to calibrate APASS data, which relies only on
the primary Landolt equatorial standards.  The comparison with Henden
and Munari secondary standards is carried out in Figure~6.  It shows
no offsets or trends with APASS $B$, $V$ photometry, and confirms 10.0 mag
as the saturation limit for the survey.  The rms of APASS data on these
secondary standards is somewhat larger than on the Landolt primary
standards used for the photometric calibration. A reason is the large
crowding affecting the low galactic fields where the Henden \& Munari 
secondary standards are located, while Landolt standards are primarely
high galactic latitude stars well isolated from neighbours.

In addition to the primary 158 photometric standards for the Sloan system,
\citet{smi02} also published a list of additional 2988 secondary
standards.  These Sloan secondary standards are {\it not} used in the APASS
calibration, so they constitute an unbiased external consistency check for
APASS photometry.  The comparison is carried out in Figure~7, and it shows
again no offset and no trend.  The larger scatter is most likely due to
the larger 24-arcsec digital aperture used by Smith et al.  Their 158 primary
standards were mostly also Landolt standards with no nearby neighbors,
but the 2988 secondary standards may not have been so carefully selected.

While there are no large sets of stars in the RAVE brightness range that
were measured by different authors in the $g'$,$r'$,$i'$ Sloan bands, the
Johnson $B$,$V$ bands have extensive coverage.  A relevant source of
data is the {\em Catalog of Mean UBV Data} (Cat-UBV for short), critically
assembled from existing literature by \citet{mer94} during the work leading
to the compilation of the Hipparcos Input Catalog.  It covers 103\,108 stars
with $U B V$ measurements spread over 693 different source papers (27\,803
stars rely on only one observation from only one source).  Figure~8
presents the comparison for the 723 field stars fainter than 10.0 mag in
both $B$ and $V$ which are in common between Cat-UBV and APASS, and for
which Cat-UBV averages over at least two distinct data sources.  Figure~9
presents a similar comparison for 1073 stars that are members of clusters.  What is
relevant in these two figures is the absence of offsets and trends in both
the magnitude zero-points and their color dependences.  The outliers in both
figures are almost entirely stars with only one observation from just two
independent sources.  The rms values in Figures~8 and 9 are somewhat larger
than for the comparisons carried out in Figures~5-7.  There are various
possible causes for this, such as ($a$) the presence of unrecognized
low-amplitude variable stars, ($b$) the mainly photoelectric origin of
Cat-UBV data (in photoelectric photometry all the light entering the
generously large aperture of the photometer - usually tens of arcsec - is
measured, thus the target together with faint field stars; while in CCD
photometry the stars are isolated from the background and measured
individually.  Perhaps not incidentally, data for members of open cluster
appear more dispersed in Figure~9 than data for field stars in Figure~8), and
($c$) the many different variants over which independent authors - whose
data were entered into Cat-UBV - tried to replicate the Johnson's $B V$
bands, Landolt's $B V$ bands being just one of such variants (a
comprehensive discussion of the realisation of the Johnson's system and the
difficulties of reproducing it with instrumentation that changed with time,
is given for example in Bessell (1990), in the books by \citet{gol74} and
\citet{str95}, and the two volumes of the Asiago Database on Photometric
Systems by \citet{mor00}, and \citet{fio03}.

The comparison between the CMC-15 and APASS $r'$ magnitudes for the 228\,805
RAVE DR4 stars in common is presented in Figure~10. There is no trend with
color, a minor systematic offset of just 0.009 mag, and a tight
concentration of points with the quartile of the distribution amounting to
only 0.018 mag. Considering the uncertainties of the photometry included in
CMC-15 mentioned above, this comparison between CMC-15 and APASS $r'$
magnitudes strongly reinforces mutual confidence in them.

  \begin{figure}[!Ht]
     \centering
     \includegraphics[width=7.5cm]{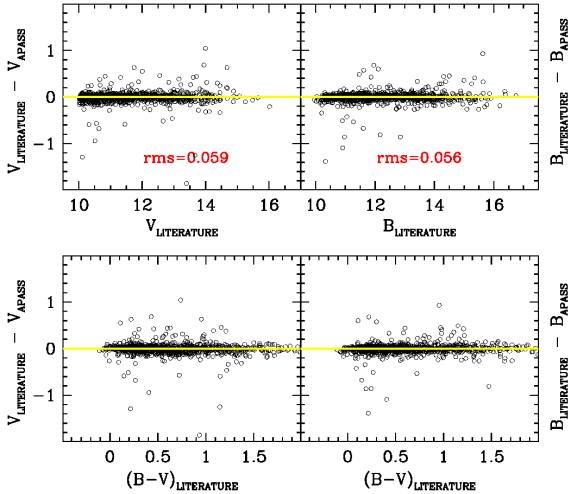}
     \caption{Comparison between literature $B$,$V$ data on stars
     belonging to open clusters as compiled by \citet{mer94}, 
     and the corresponding APASS values.\label{fig9}}
  \end{figure}

  \begin{figure}[!Ht]
     \centering
     \includegraphics[width=7.5cm]{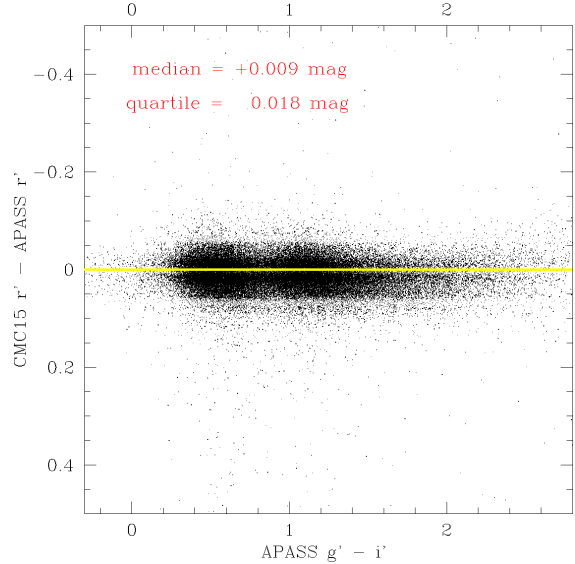}
     \caption{Comparison between the Carlsberg Meridian Catalog
     (CMC-15) and APASS $r'$ magnitudes for the 262\,972 RAVE stars
     in common.\label{fig10}}
  \end{figure}

\subsection{Distribution in magnitude and over the HR diagram of RAVE stars}

The distribution in APASS and 2MASS magnitudes for RAVE DR4 stars is shown in
Figure~11, and it is characterized by the following median values (which are not
affected by potential saturation problems at the very bright end of the
distributions):

\begin{eqnarray}
{\rm med}(B) &=&12.654 \nonumber \\
{\rm med}(g')&=&12.126 \nonumber \\
{\rm med}(V) &=&11.676 \nonumber \\
{\rm med}(r')&=&11.362           \\
{\rm med}(i')&=&11.033 \nonumber \\
{\rm med}(J) &=& 9.802 \nonumber \\
{\rm med}(H) &=& 9.330 \nonumber \\
{\rm med}(K) &=& 9.214 \nonumber 
\end{eqnarray}

The stars for the RAVE input catalog were selected on the basis of their
$I$-band magnitude (coming from DENIS, or SuperCOSMOS catalogs, or
extrapolated from other sources like Tycho-2), generally within the
boundaries $9 \leq I \leq 12$.  To avoid problems from fiber cross-talk,
the targets for any given RAVE observation should span the strictest possible
range in magnitude that would still provide enough targets within the
$6^\circ$ field of view to cover all available fibers.  This led to grouping
the targets in the RAVE input catalog over four distinct ranges in $I$-band
magnitude (nominally $9.0-10.0$, $10.0-10.75$, $10.75-11.5$, $11.5-12.0$). 
The corresponding tailing of the actual observations is apparent in the
segmented distribution affecting the $i'$ panel of Figure~13.  The presence
of segmentation in other bands declines with increasing wavelength
separation from the $i'$ band. It is noteworthy that the distribution in
$B$ band is closely matched by a Gaussian (overplotted) centered on
$B =12.62$ and with $\sigma =1.11$~mag.

  \begin{figure*}
     \centering
     \includegraphics[height=16.0cm,angle=270]{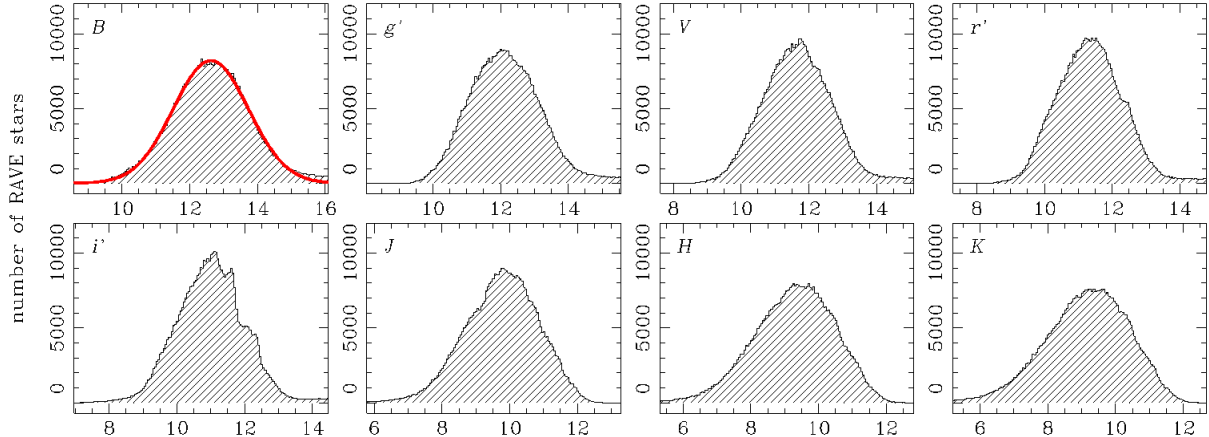}
     \caption{Distribution ($\Delta$mag=0.05 bins) in APASS
     $B V g' r' i'$ and 2MASS $J H K$ magnitudes for RAVE DR4 stars.
     The segmented distribution in $i'$ band is the result of RAVE stars
     being selected from different input catalogs and the observations being
     executed according to $I$ band brightness intervals.  A Gaussian is
     over-plotted for reference to the $B$-band distribution.\label{fig11}}
  \end{figure*}

The distribution of RAVE stars over the HR diagram is shown in Figure 12,
and is dominated by closer, hotter and higher galactic latitude Main
Sequence/Turn-Off stars (the cloud of points centered at
$B-V=0.6$, $\log g=4.4$), and by more distant, cooler and lower galactic
latitude RGB and Red Clump stars (the cloud of points centered at
$B-V=1.1$, $\log g=2.3$).

\section{Fitting to a synthetic photometric grid}

To extract information about the observed stars, we performed a $\chi^2$ fit
of APASS+2MASS photometry against a library of purposefully-built synthetic
photometry.  The source stellar energy distributions are the atmospheric
flux models of \citet{cas03}.  Their model atmospheres are the
basis of the \citet{mun05} synthetic spectral atlas, which has been
used to derive atmospheric parameters from spectra in RAVE Data Releases 2
and 3.  These models have been extensively used in the literature, including
the recent work by \citet{bes12} on revised passbands and zero
points for UBVRI, Hipparcos, and Tycho photometry.

The atmospheric flux models of \citet{cas03} cover, at regular
steps, a wide portion of the atmospheric parameter space along the effective
temperature ($T_{\rm eff}$), surface gravity ($\log g$), and metallicity
([M/H]).  Two values of the enhancement in $\alpha$-elements ([$\alpha$/M])
are considered (0.0 and +0.4), while the micro-turbulence is constant at 2
km~sec$^{-1}$.  To increase the density of grid points in the range of
temperatures covered by the vast majority of RAVE stars, the original
atmospheric flux models for $T_{\rm eff} \leq 9750$~K have been parabolic
interpolated to $\Delta T_{\rm eff} = 50$~K, $\Delta \log g =0.25$~dex,
$\Delta$[M/H]=0.1, and a third value (+0.2) for [$\alpha$/M] has been
derived from linear interpolation of source values (0.0 and +0.4). 
For hotter temperatures ($T_{\rm eff} > 9750$), where less than 1\%\ of all
RAVE stars are found, only one interpolated mid-point has been added between
the original \citet{cas03} grid points.  The atmospheric flux
models have then been reddened according to 64 different values of
$E_{B-V}$, from 0.0 to 2.0, with a finer step at lower reddenings
($\Delta E_{B-V} =0.005$) that progressively widens with increasing
extinction (last two points at $E_{B-V} =1.75$ and 2.00).  Considering the
high galactic altitude of RAVE target stars, for the interstellar reddening
law we adopted the standard $R_V = A_V / E_{B-V} =3.1$ law as formulated by
\citet{fit99}.  A total of 13 million interpolated and reddened
atmospheric flux models were used to build the photometric synthetic
library.

The magnitudes in the eight bands of interest (Landolt $B$,$V$, Sloan
$g'$,$r'$,$i'$, 2MASS $J$,$H$,$K$) were obtained for all input
atmospheric flux models by direct integration:
\begin{equation}
{\rm mag} = -2.5 \log \frac{\int 10^{(-0.4R_\lambda E_{B-V})}I(\lambda) 
            S(\lambda) d\lambda}{\int S(\lambda) d\lambda} + {\rm const}
\end{equation}
where I($\lambda$) is the flux distribution of the atmospheric model, and
S($\lambda$) is the transmission profile of the given photometric band.  The
zero-points were set to match the observed values for the primary
photometric standard of the Sloan system BD+17.4708\footnote{Bohlin and
Gilliland (2004b) provided an accurate absolute flux distribution of this
star based on HST data}, for which there are
several published determinations of the atmospheric parameters.  The
corresponding mean values are $T_{\rm eff} =5980$~K, $\log g =3.85$ and
[M/H]$= -1.72$.  An [$\alpha$/Fe]=+0.4 atmospheric flux model for these
parameters was obtained by parabolic interpolation from \citet{cas03}, 
and it was reddened by $E_{B-V} =0.010$ following \citet{ram06}. 
The zero points in Eq.(2) then follow from the requirement
that the computed values match the observed values $B =9.911$, $V=9.467$,
$g'=9.640$, $r'=9.350$, $i'=9.250$ \citep[from][]{fuk96,smi02}, 
and $J=8.435$, $H=8.108$, $K=8.075$ (2MASS catalog) for the
BD+17.4708 standard.

  \begin{figure}[!Ht]
     \centering
     \includegraphics[height=7.5cm]{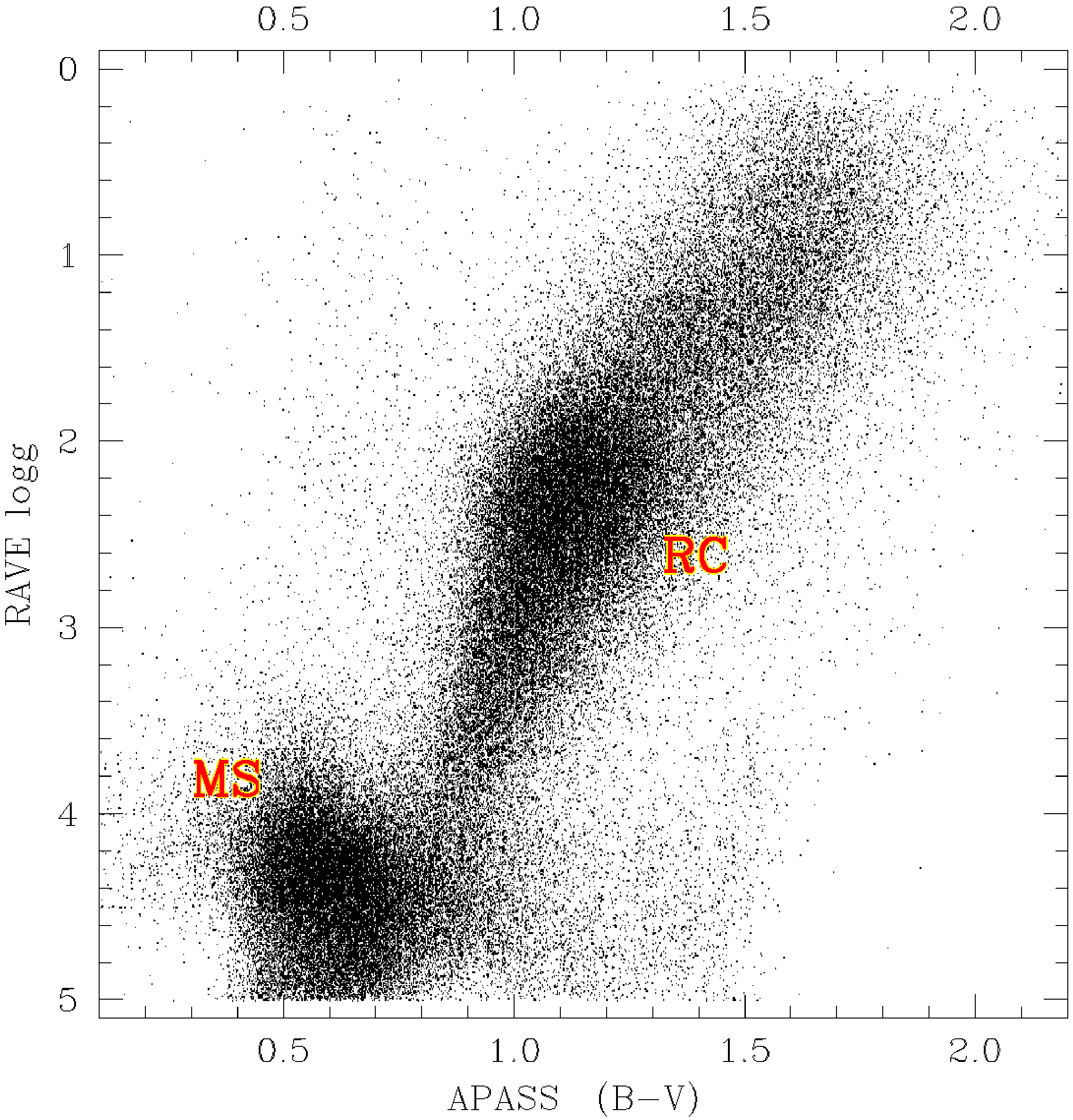}
     \includegraphics[height=7.5cm]{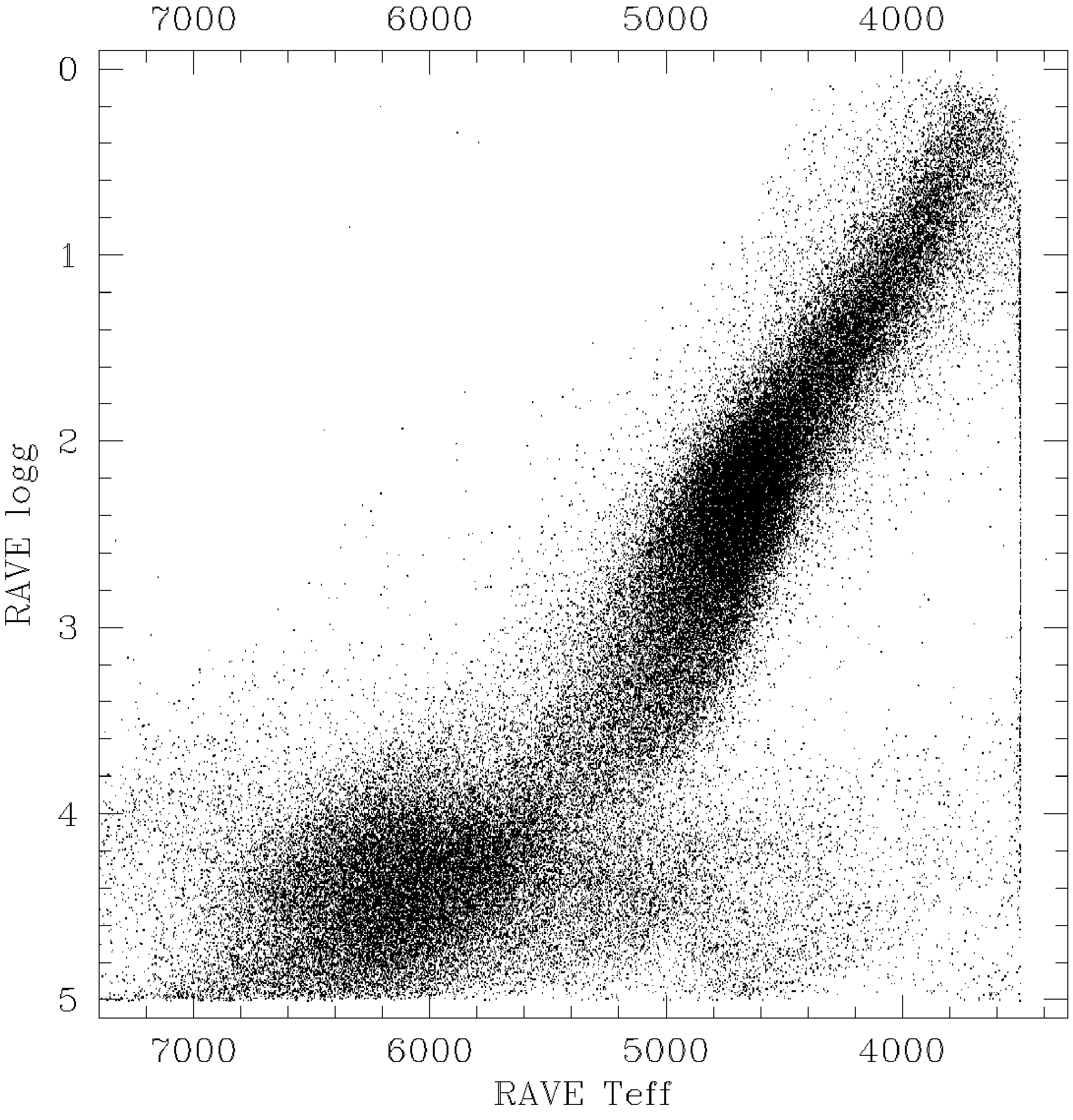}
     \caption{Distribution of RAVE stars on the spectroscopic $T_{\rm eff}$
      vs $\log g$ plane (bottom), and replacing $T_{\rm eff}$ with APASS
      $B-V$ (top). The same stars are plotted on both panels.\label{fig12}}
  \end{figure}
  
To derive the best matching atmospheric parameters and reddening, the
observed APASS $+$ 2MASS $B$,$V$,$g'$,$r'$,$i'$,$J$,$H$,$K$ magnitudes
(mag$^\ast$) were $\chi^2$ fitted to the synthetic photometric grid:
\begin{equation}
     \chi^2 = \sum_{i} \frac{({\rm mag^\ast_{i}} - {\rm mag_{i})^2}}{\epsilon^{2}_{i}}
\end{equation}
where $i$ sums over the eight photometric bands and $\epsilon_{i}$ are the
individual errors quoted in APASS and 2MASS catalogs.  Extensive simulations
on synthetic data and analysis of real data on RAVE stars, shows that the
great density of points in the matched synthetic photometric library
produces $\chi^2$ minima which are symmetric around their core.  To derive
the best fit values, we therefore averaged over the five deepest points
within the $\chi^2$ minimum.

\section{Temperature-Reddening degeneracy}

The observed spectral energy distribution of a normal star is primarily
dominated by the combined effect of temperature and reddening.  The two
cannot be fully disentangled from optical+IR photometric data alone. This is
demonstrated in Figure~13, where the results of a Monte Carlo simulation are
presented.  We took the photometry of one atmospheric flux model (from the
13\,244\,000 entries of the synthetic photometric reference grid described
in the previous section), Monte Carlo perturbed it according to a Gaussian
distribution of errors (with $\sigma$ given by the values in Eq.1), and
$\chi^2$ fitted it back to the photometric reference grid.  Figure~13
presents the distribution of the fitting results, on three separate $T_{\rm
eff} - E_{B-V}$ planes to highlight the effect of alpha-enhancement,
surface gravity and metallicity, while the last panel on each row shows the
surface gravity vs.  metallicity correlation.

In the Monte Carlo simulations of Figure~13, we considered two cases, a main
sequence/turn-off star ($T_{\rm eff} =5800$~K, $\log g =4.0$,
[$\alpha$/Fe]=0) and a red clump star ($T_{\rm eff} =4650$~K, $\log g
=2.75$, [$\alpha$/Fe]=0), either metal rich ([M/H]=+0.2) and metal poor
([M/H]=$-$0.6), and suffering from $E_{B-V}=0.06$ or $E_{B-V}=0.34$
reddening, for a total of eight different combinations.  The results of
Figure~13 say that broad-band photometry, even if widely distributed in
wavelength over the optical and infrared ranges, and of state-of-the-art
accuracy, cannot accurately disentangle temperature from reddening, the two
are intimately degenerate, and that the possibility to constrain
gravity and metallicity is, at best, marginal.  This is not surprising,
since we have at least five variables ($E_{B-V}$, $T_{\rm eff}$, $\log g$,
[M/H], [$\alpha$/Fe]) and only eight data points to constrain them (whereas
spectra can count on hundreds of absorption lines as inputs).  Here we
considered the ideal case of $\chi^2$ fitting synthetic vs.  synthetic data. 
In the real world, with real stars, there are several additional variable
factors to further confuse the picture: ($i$) a fraction of the real stars
are binaries, merging the light of two or more stars of different properties
\citep[including the possibility of even different circumstellar reddening,
e.g.][]{all83}, ($ii$) the cosmic dispersion of real stars (for ex.  in
chemical partition or micro-turbulence) cannot be accounted for by any
synthetic photometric grid, ($iii$) peculiarities like chromospheric
activity (a common feature of young main sequence stars), winds and
circumstellar material (frequent in cool giants, yellow supergiants or the
hotter stars), or even the latitudinal dependence of colors on fast rotating
stars \citep[e.g.][]{gol74,boh04}, and the ($iv$) differences in shapes
that the wavelength dependence of interstellar reddening can take over the
wavelength range of interest \citep[from $R_V$=2.1 to $R_V$=5.0,
e.g.][]{car89,fit99}.  These uncertainties are added to the known
difficulties of synthetic spectra to accurately reproduce observed ones,
expecially at the borders of the explored parameter space.

  \begin{figure*}[!Ht]
     \centering
     \includegraphics[width=16cm]{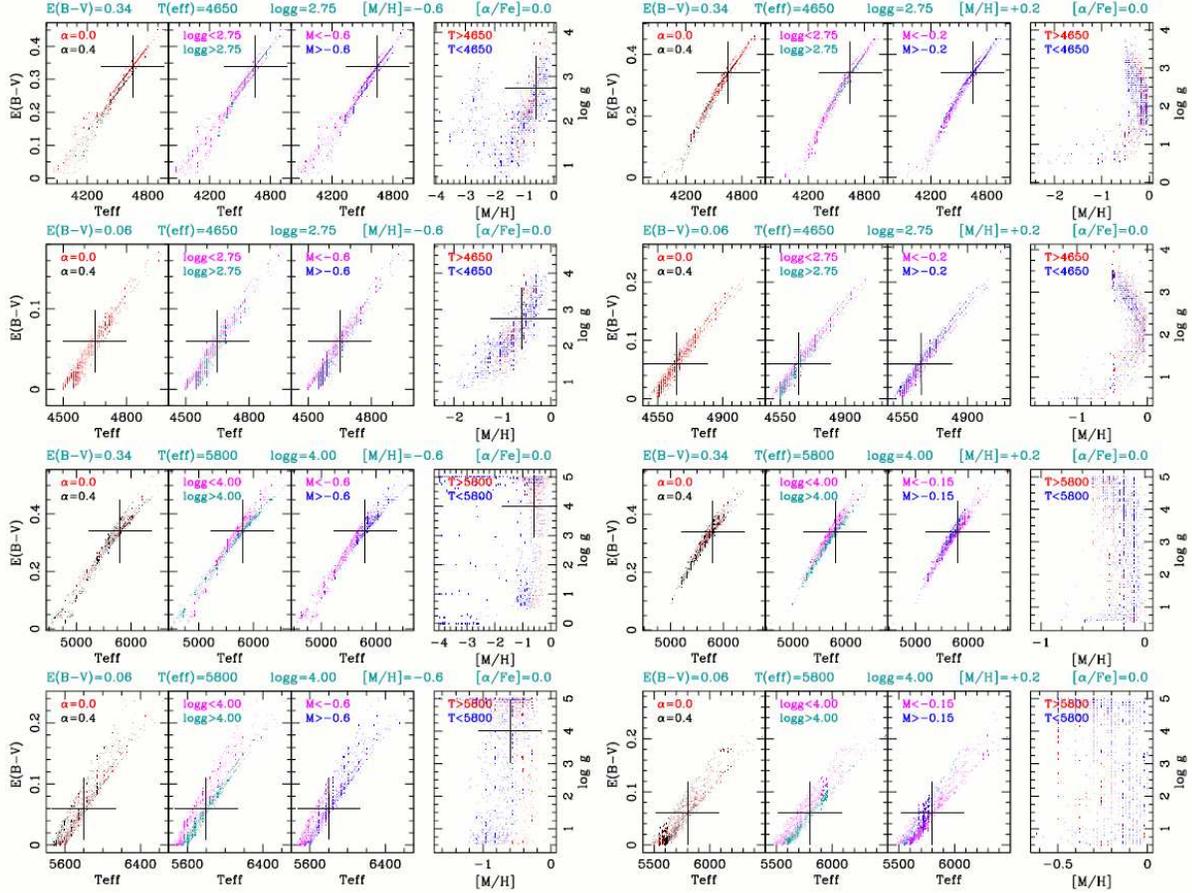}
     \caption{Degeneracy between reddening and temperature on multi-band
     photometry of stars.  The $B V g' r' i' J H K$ magnitudes of
     eight synthetic atmospheric flux distributions (for a Red Clump and a
     Main Sequence star, either metal rich or poor, and either significantly
     or marginally reddened) are Monte Carlo perturbed according to APASS
     and 2MASS typical errors for RAVE stars (cf Eq.~1 and Figure~3), and
     then subjected to $\chi^2$ fit to the synthetic photometric library
     (see sect.  4 for details).  The crosses mark the position of the
     reference unperturbed model.\label{fig13}}
  \end{figure*}

Below in this paper, we will provide for RAVE stars several different values of
temperature and reddening. In addition to those derived by fitting
APASS+2MASS data simultaneously for both of them, we'll also fix the reddening
and derive the temperature and conversely fix the temperature and derive the
reddening, so to provide the reader with a range of options.

\section{Dust model parameterized on standard stars}

To derive the temperature once the reddening is fixed, we introduce a simple
reddening model.  We assume that the low reddening affecting the high
Galactic latitude RAVE stars originates from an homogeneous slab of dust,
extending symmetrically for $d$ parsecs above and below the Galactic plane,
with the Sun positioned on the Galactic plane.  In this model, sketched in
Figure~14, the amount of reddening is directly proportional to the length
traversed by the line-of-sight within the dust slab (indicated by the
thicker portions of the lines-of-sight in Figure~14).  The model is
completely defined by the thickness of the slab ($d$), the total reddening
at the pole ($E^{poles}_{B-V}$, the same value for the north and south
poles), and the reddening law.  We take the reddening law to be the standard
$R_V$=3.1, valid for the diffuse interstellar medium \citep{sav79}. 
Investigation of the actual 3D Galactic reddening map as inferred from
APASS+2MASS data for RAVE stars will be carried out elsewhere in this series
of papers.  It is worth noticing that typical RAVE stars are at high
galactic latitudes and at a significant distance from the Sun.  Thus most of
them lie outside the dust slab, and the distribution of dust within the slab
is not important so that we can characterize it with just two numbers, its
thickness and its extinction at the poles. Also the presence of the
local interstellar bubble around the Sun (e.g.  Abt 2011 and references
therein) is of little concern here, all RAVE and calibration stars below
considered lying similarly well beyond it.

  \begin{figure}[!Ht]
     \centering
     \includegraphics[width=7.5cm]{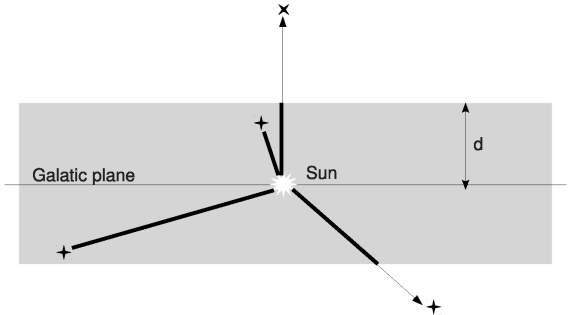}
     \caption{Outline of the simple Galactic dust distribution model
     described in sect.5 and used to statistically account for the
     interstellar extinction affecting high galactic latitude RAVE stars.
     \label{fig14}}
  \end{figure}

To derive the amount of total reddening at the poles and constrain the
thickness of the dust slab, we considered two samples of stars,
characterized by having both accurate atmospheric parameters (derived from
high resolution, high S/N spectroscopy) and accurate APASS and 2MASS
photometry.  A first sample comes from \citet{ruc10,ruc11a,ruc11b}, and a
second sample is composed of stars from the PASTEL database \citet{sou10}. 
Distances to these stars were obtained from isochrone fitting as
described by \citet{zwi10}.  These distances are homogeneous with
those to RAVE stars computed in the same way (see below).  The sample of
Ruchti+PASTEL stars with unsaturated APASS+2MASS magnitudes is composed by
242 stars.  They are distributed similarly to RAVE stars in terms of
temperature and surface gravity: those hotter than 5300~K are generally main
sequence objects, while giants dominates at lower temperatures.  In both the
RAVE survey and the Ruchti+PASTEL sample the number of giants is larger than
the main sequence stars.  They however differ significantly in metallicity:
the median metallicity of Ruchti+PASTEL main sequence stars is $-$0.85, and
$-$1.41 is that of giants, those of RAVE stars are $-$0.17 and $-$0.43,
respectively.  When unsaturated APASS photometry will be available for stars
up to 7.5~mag, it will be possible to include a larger proportion of
standard stars of higher metallicity.

The APASS+2MASS data for the stars in the Ruchti+PASTEL sample were
subjected to the $\chi^2$ fitting described in sect.3.  A dense grid of
values for the thickness of the slab ($d$) and the total reddening at the
pole ($E^{poles}_{B-V}$) was explored.  The photometric temperature derived from
the $\chi^2$ fitting was then compared to the spectroscopic values, and the
$\Delta T_{\rm eff}$ difference evaluated.  The
values of $d$ and $E^{poles}_{B-V}$ that best represent the Galactic
extinction are those that tend to null the trend of $\Delta T_{\rm eff}$
with $T_{\rm eff}$, and that minimize both the median value and the rms of
$\Delta T_{\rm eff}$.  

  \begin{table*}[!Ht]
     \centering
     \caption{APASS photometric and astrometric data of RAVE DR4 stars, and
     their $T_{\rm eff}$ and $E_{B-V}$ derived - under different options -
     via $\chi^2$ fitting to an extensive synthetic photometric library. See
     sect. 6 for a detailed description of the content of the thirty-six
     columns (the table is published in its entirety in the electronic
     edition of the journal.  A single line is shown here for guidance
     regarding its content).\label{tab}}
     \includegraphics[width=16cm]{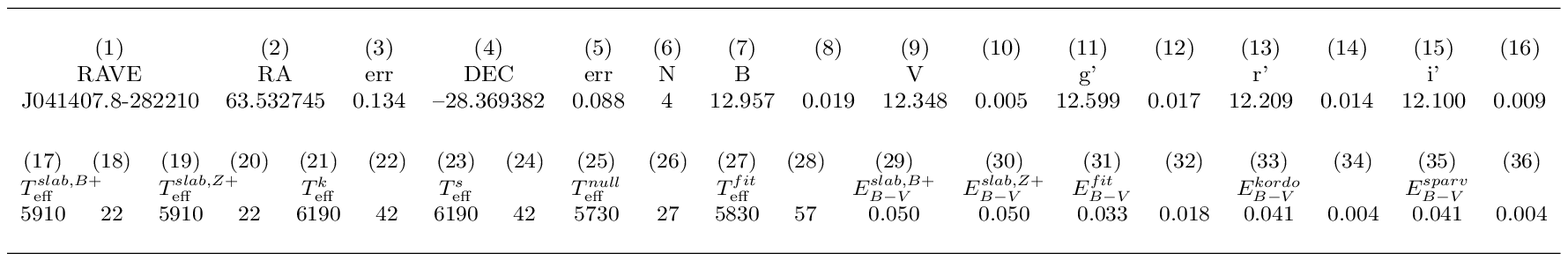}     
  \end{table*}

 \begin{figure*}[!Ht]
     \centering
     \includegraphics[width=13cm]{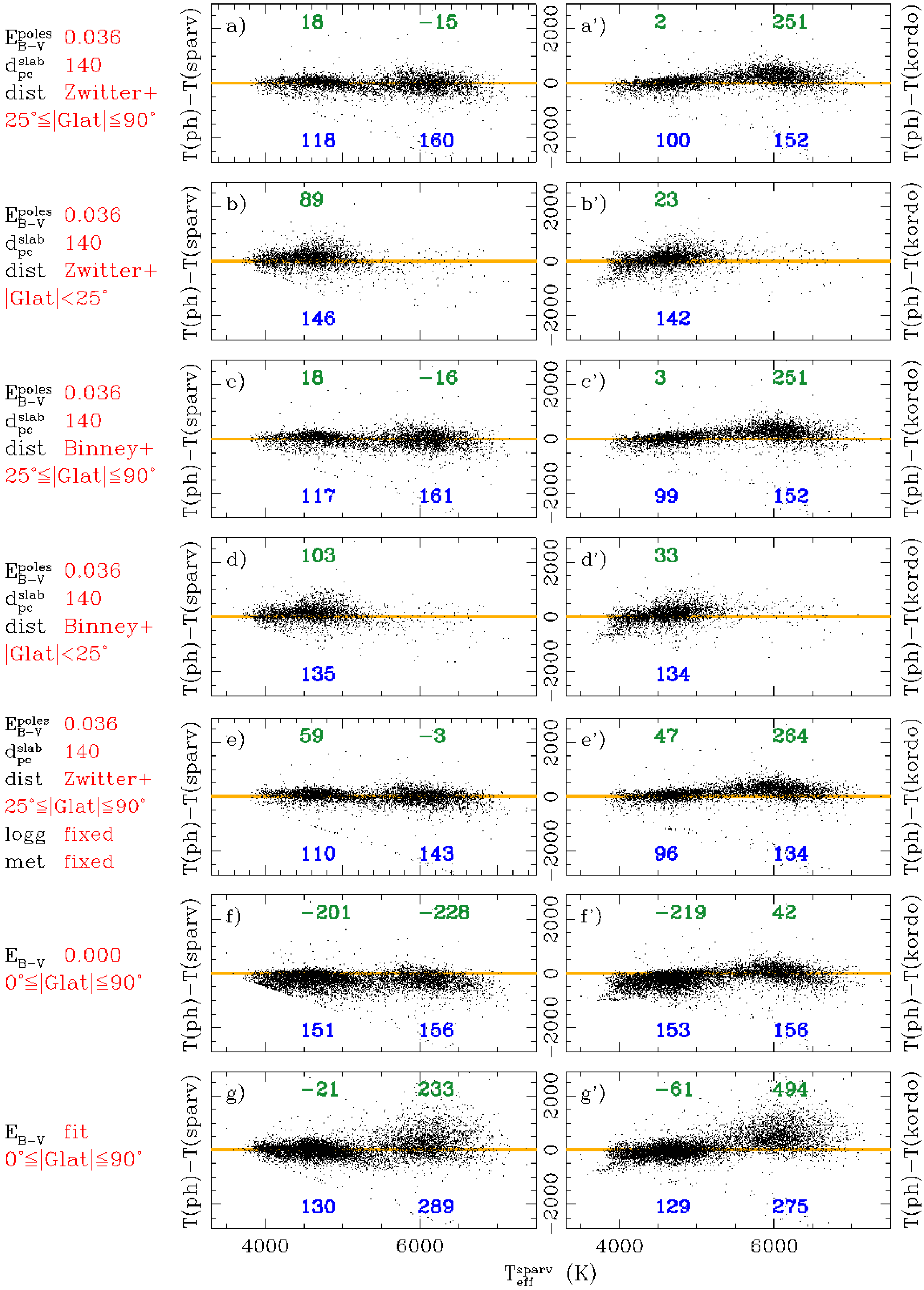}
     \caption{Difference between RAVE DR4 spectroscopic and APASS+2MASS photometric
     temperatures from Table~1 (cf. sect.7 for explanation and discussion).
     In each panel, the upper two numbers (in green) are the median value of
     the ordinate and the lower two (in blue) are the quartile values. They
     are listed separately for stars hotter and cooler than 5250~K, an
     handy value dividing the RBG+RC stars (on the cooler side) from the 
     MS/TO stars (on the hotter side).\label{fig15}} 
  \end{figure*}

A value of $E^{poles}_{B-V} =0.036 \pm 0.003$ and $d =140 \pm 30$~pc are
found.  It is worth noting that the $\chi^2$ fitting assumes the reddening
affecting the Sloan primary photometric standard BD+17.4708 to be
$E_{B-V} =0.010$ following \citet{ram06}.  Any deviation from
$E_{B-V} =0.010$ of the reddening affecting BD+17.4708 should be added to
$E^{poles}_{B-V}$.  In the existing literature, various methods have been
used to derive $E^{poles}_{B-V}$ (stellar counts and colors, galaxy counts
and colors, interstellar polarization).  A probably incomplete list of
available literature provides the following determinations (in chronological
order): $E^{poles}_{B-V}$=0.023 by \citet{str67}, 0.029
by \citet{bon71}, 0.018 by \citet{phi73}, 0.054 by \citet{hol74},
0.03 by \citet{app75}, 0.048 by \citet{hei76}, 0.057 by \citet{knu77},
0.00 by \citet{bur78}, 0.060 by \citet{lyn79}, 0.024 by \citet{alb80}, 
0.04 by \citet{nic82}, 0.048 by \citet{dev83}, 
0.00 by \citet{mcf83}, 0.050 by \citet{egg95}, 
0.055 by \citet{ber01}, and $>$0.043 by \citet{ber04}. 
Their mean value is $E^{poles}_{B-V} =0.036$ ($\sigma =0.019$, and 0.005 as
the error of the mean), the same value we found from APASS+2MASS data of
Ruchti+PASTEL stars, which reinforces the confidence on the results of the
present paper. The value of 140 pc we found for $d$ well matches the scale
height of the youngest stellar population of the Galaxy with which dust is
intimately related.

Future extension of the APASS survey to cover stars up to 7.5 mag brightness
will greatly increase the number of nearby spectroscopic standard stars,
especially among the main sequence stars, and this will provide a stronger
handle to fix $E^{poles}_{B-V}$ and $d$, as well as for investigating
differences in the total extinction at the north and south Galactic poles.

 \begin{figure*}[!Ht]
     \centering
     \includegraphics[width=12cm]{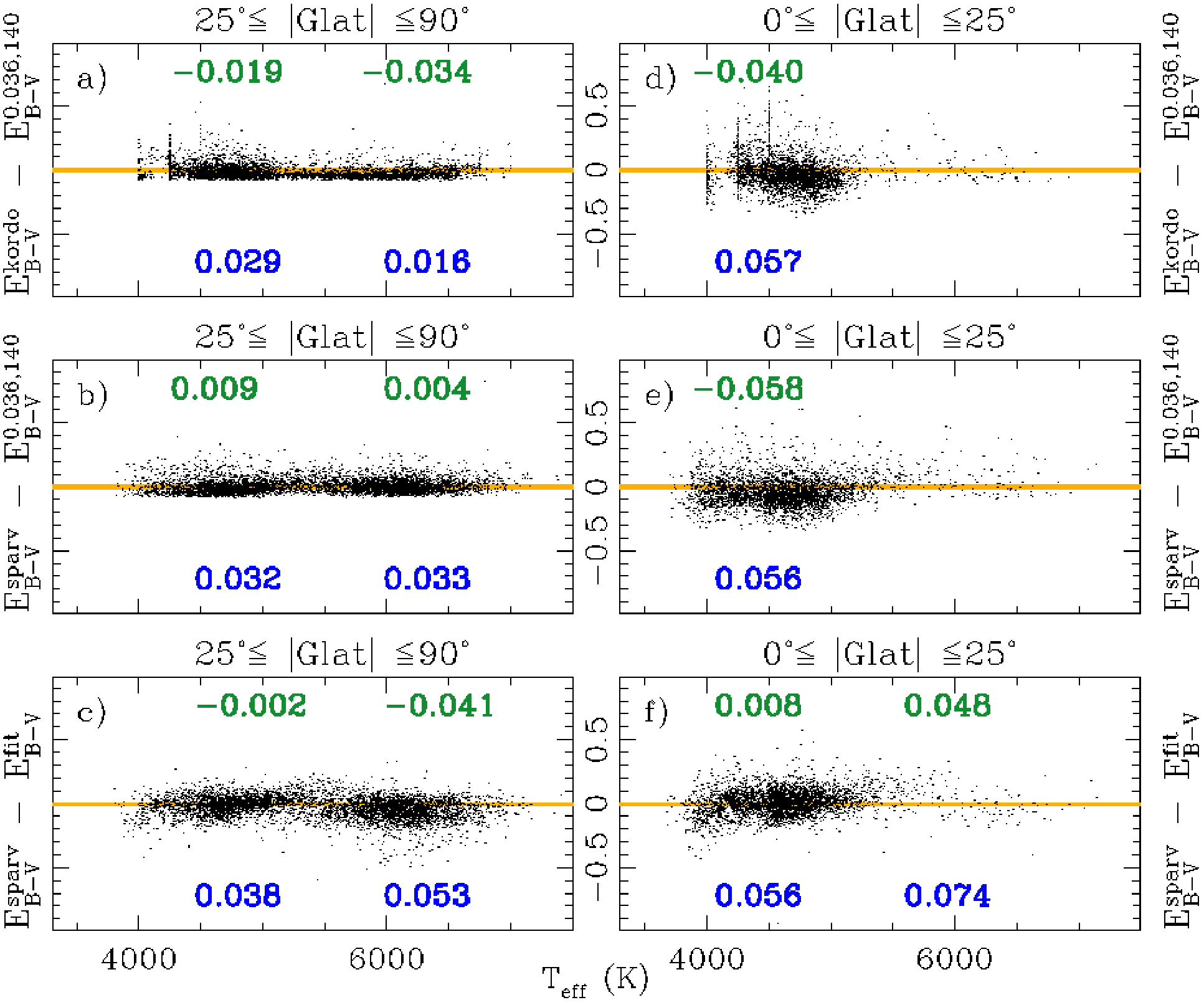}
     \caption{The upper two rows compare the slab reddening (cf. sect. 5) to
     the reddening derived from $\chi^2$ fit to APASS+2MASS photometry
     fixing the atmospheric parameters to {\em kordo} and {\em sparv} values
     from RAVE DR4. The last row compares the reddening derived from
     simultaneous fitting to all parameters to that obtained by fixing the
     atmospheric parameters to {\em sparv} values from RAVE DR4. The panels
     at left focus on RAVE stars at $|b| \geq 25 ^\circ$, those at right to
     RAVE stars at $|b| \leq 25^\circ$.\label{fig16}}
  \end{figure*}

\section{Results}

The photometric temperature and reddening of RAVE stars have been computed
according to the $\chi^2$ procedure described in sect.3 and the results are
given in Table~1. The full table is available electronic only via CDS, the
electronic edition of this journal and the RAVE web site. The electronic
version of the table will be kept updated in response to future RAVE and
APASS data releases.

In addition to provide APASS astrometric positions and magnitudes (with
their respective errors) for RAVE DR4 stars, Table~1 lists a series of
values for the temperature and reddening computed under different
conditions, outlined below.  Table~1 lists results for 377\,379 RAVE
spectra (336\,644 stars) and does not include the RAVE stars that have been
classified as peculiar by \citet{mat12}, which will be
studied separately and in conjunction with epoch APASS photometry.  For
uniformity reasons, Table~1 also does not include the few RAVE stars for
which data are missing in one or more of the $B V g' r' i' J H K$
photometric bands.  By {\em slab} reddening in Table~1, we mean the
reddening produced by the line of sight to the given star traversing an
homogeneous slab of interstellar dust, extending by $d$=140 pc on either
sides of the Galactic plane (with the Sun centered on it), following the
standard $R_V=3.1$ law and causing a total $E^{poles}_{B-V} =0.036$~mag at
both Galactic poles.  It is completely defined by the Galactic latitude and
the distance to the given star.  Finally, RAVE DR4 computed the atmospheric
parameters ($T_{\rm eff}$, $\log g$ and [M/H]) from observed spectra
following two different approaches, named {\em kordo} and {\em sparv}, which
are described in \citet{kor13} (where preference is given to {\em kordo}
values) and that are both used in this paper. 

The columns of Table~1 provide the following data. Column: \\
(1) RAVE name;\\
(2) APASS right ascension (equinox J2000, epoch 2013);\\
(3) its error (in arcsec);\\
(4) APASS declination (equinox J2000, epoch 2013);\\
(5) its error (in arcsec);\\
(6) number N of independent photometric nights during which the star was
    measured by APASS;\\
(7) $B$ mag;\\
(8) the $\sigma$ of the N (column 6) independent measurements;\\
(9) $V$ mag;\\
(10) the $\sigma$ of the N (column 6) independent measurements;\\
(11) $g'$ mag;\\
(12) the $\sigma$ of the N (column 6) independent measurements;\\
(13) $r'$ mag;\\
(14) the $\sigma$ of the N (column 6) independent measurements;\\
(15) $i'$ mag;\\
(16) the $\sigma$ of the N (column 6) independent measurements;\\
(17) $T_{\rm eff}$ (K) computed by assuming the {\em slab} reddening
     corresponding to \citet{bin14} distance;\\
(18) its uncertainty (K) expressed as the rms of the five
     deepest points within the $\chi^2$ minimum;\\
(19) $T_{\rm eff}$ (K) computed by assuming the {\em slab} reddening
     corresponding to \citet{zwi10} distance;\\
(20) its uncertainty (K) expressed as the rms of the five
     deepest points within the $\chi^2$ minimum;\\
(21) $T_{\rm eff}$ (K) computed by assuming the {\em slab} reddening
     corresponding to \citet{zwi10} distance, and fixing the values of
     $\log g$ and [M/H] to ``{\em kordo}'' values from RAVE DR4;\\
(22) its uncertainty (K) expressed as the rms of the five
     deepest points within the $\chi^2$ minimum;\\
(23) $T_{\rm eff}$ (K) computed by assuming the {\em slab} reddening
     corresponding to \citet{zwi10} distance, and fixing the values of
     $\log g$ and [M/H] to ``{\em sparv}'' values from RAVE DR4;\\
(24) its uncertainty (K) expressed as the rms of the five
     deepest points within the $\chi^2$ minimum;\\
(25) $T_{\rm eff}$ (K) computed by assuming null reddening;\\
(26) its uncertainty (K) expressed as the rms of the five
     deepest points within the $\chi^2$ minimum;\\
(27) $T_{\rm eff}$ (K) computed by letting the
     $\chi^2$ to simultaneously fit $E_{B-V}$;\\
(28) its uncertainty (K) expressed as the rms of the five
     deepest points within the $\chi^2$ minimum;\\
(29) {\em slab} $E_{B-V}$ reddening for \citet{bin14} distance;\\
(30) {\em slab} $E_{B-V}$ reddening for \citet{zwi10} distance;\\
(31) $E_{B-V}$ (mag) computed by letting the $\chi^2$ to simultaneously
     fit $T_{\rm eff}$;\\
(32) its uncertainty (mag) expressed as the rms of the five deepest
     points within the $\chi^2$ minimum;\\
(33) $E_{B-V}$ (mag) computed by fixing $T_{\rm eff}$,
     $\log g$ and [M/H] to ``{\em kordo}'' values from RAVE DR4;\\
(34) its uncertainty (mag) expressed as the rms of the five deepest
     points within the $\chi^2$ minimum;\\
(35) $E_{B-V}$ (mag) computed by fixing $T_{\rm eff}$, $\log g$ and [M/H] to
     ``{\em sparv}'' values from RAVE DR4;\\
(36) its uncertainty (mag) expressed as the rms of $E_{B-V}$ of the five deepest
     points within the $\chi^2$ minimum.

Some remarks to Table~1. The uncertainties of APASS photometry reported in
columns 8, 10, 12, 14, and 16 are the $\sigma$ of the N independent
measurements in that band, so that the error ($\epsilon$) on the reported
magnitude should be computed as $\sigma / \sqrt{N}$. When a "-" sign
precedes the value of $\sigma$, it means that less than N observations have
been obtained in that specific band (in performing the $\chi^2$ procedure
above described, we have assumed $\epsilon_i = \sigma_i$ in such a case,
equivalent to assume that only one observation was obtained in that band).
When neither a valid \citet{zwi10}, nor a \citet{bin14} 
distance exists for a given RAVE star (as for the stars included in the
first RAVE Data Release), its distance is set to 0 and consequently the
corresponding {\em slab} $E_{B-V}$ reddening is null. The error in right
ascension of APASS positions, expressed in arcsec, already takes into
account the declination projection factor.

 \begin{figure*}[!Ht]
     \centering
     \includegraphics[width=15cm]{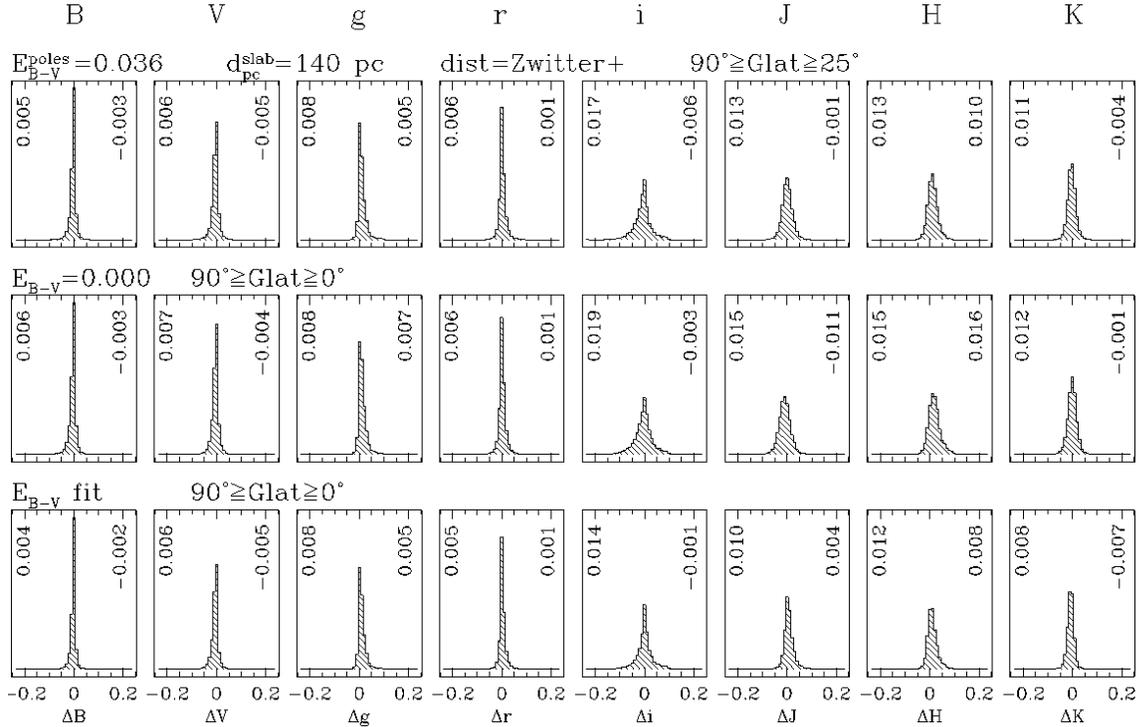}
     \caption{Distribution of the $\chi^2$ fitting residuals (mag) for, top to
     bottom, panels $a$, $f$ and $g$ of Figure~16. In each panel, the
     vertical number to the left in the rms (mag) and that to the right the
     median (mag).\label{fig17}}
  \end{figure*}

 \begin{figure*}[!Ht]
     \centering
     \includegraphics[width=14cm]{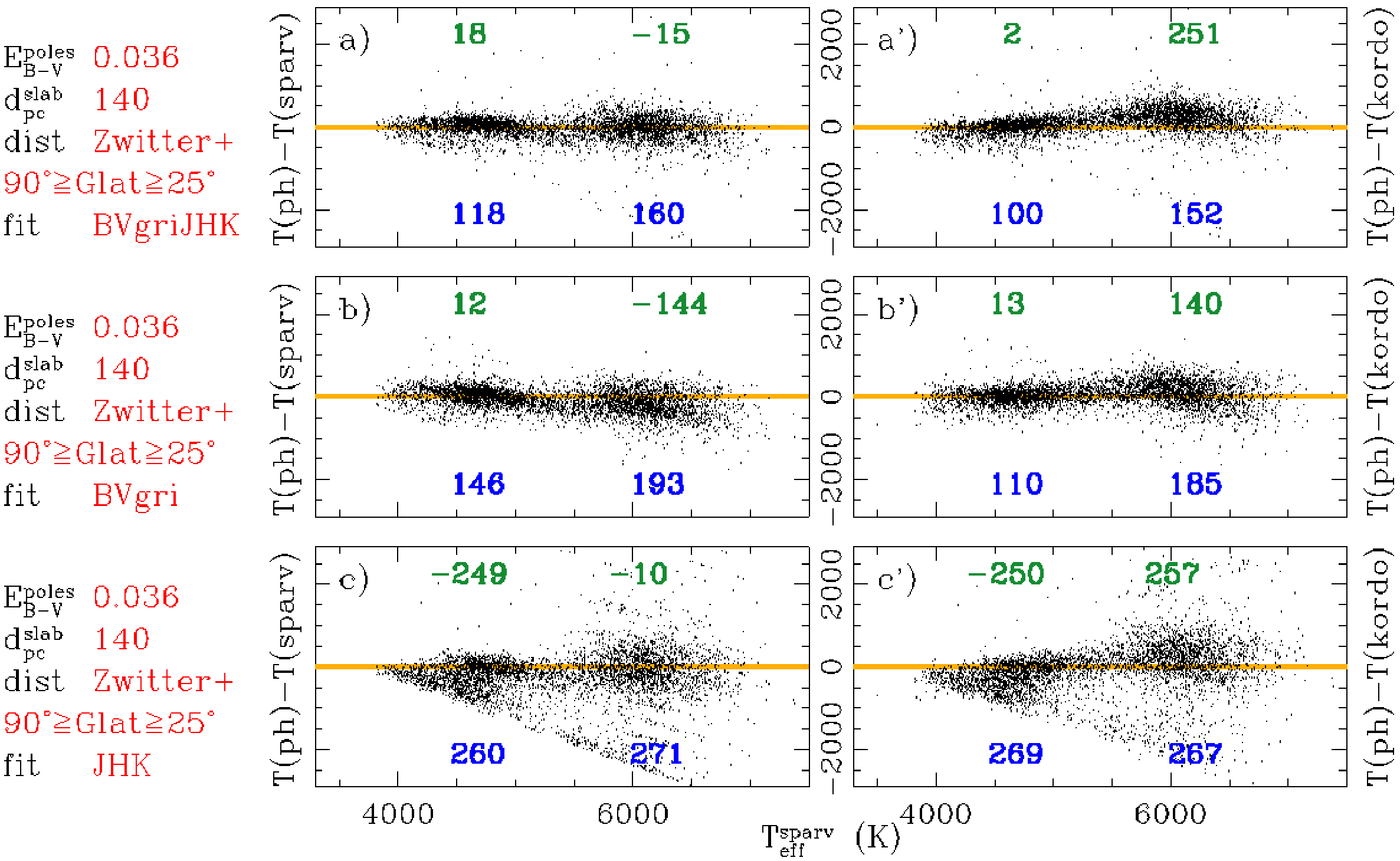}
     \caption{The effect of ignoring 2MASS (middle panels) or APASS data
     (bottom panels) in the $\chi^2$ fitting of RAVE DR4 stars. Panels $a$,$a'$ 
     are copied from Figure~16 for comparison purposes.\label{fig18}}
  \end{figure*}

When 0.0 is listed as the rms of the five deepest points within the $\chi^2$
minimum for either $T_{\rm eff}$ or $E_{B-V}$, it means these points differ
only in the other fitted parameters (e.g. $\log g$, [M/H], [$\alpha$/Fe]). 
Unless otherwise indicated (see above), in computing the columns of Table~1
$\log g$, [M/H], and [$\alpha$/Fe] are let free to be fitted.  Given the low
sensitivity of broad band photometry to parameters other than $T_{\rm eff}$
and $E_{B-V}$, we refrain from listing the fitting results for $\log g$,
[M/H] and [$\alpha$/Fe]. We plan to reconsider this point at a later time
when (1) final APASS products will be released that will further reduce its
already excellent photometric errors, (2) extension of APASS toward brighter
saturation limits will enable to test the sensitivity to $\log g$, [M/H] and
[$\alpha$/Fe] on a much larger set of well studied standard stars than
currently available (the vast majority of them are brighter than 10 mag).

\section{Discussion}

The temperature of RAVE stars obtained from $B V g' r' i' J H K$
photometry is compared in Figure~15 to spectroscopic temperature from
RAVE DR4. Photometric and spectroscopic temperatures agree well. 

In the top four rows of Figure~15, this comparison is carried out separately
for stars above and below 25$^\circ$ in galactic latitude, which was the
original limit of the RAVE survey to avoid the regions closer to the
galactic plane where the reddening becomes relevant and increasingly patchy. 
The comparison is carried out separately for \citet{zwi10} and
\citet{bin14} distances to RAVE stars, and for the two different
types of spectroscopic temperatures $T_{\rm eff}^{\rm sparv}$ and $T_{\rm
eff}^{\rm kordo}$ derived in RAVE DR4.

To filter out outliers and lower quality data, the comparison between
photometric and spectroscopic temperatures in Figure~15 is carried out only
for stars with best RAVE and APASS data.  Only stars with 4 or more
independent APASS photometric observations and fainter than the 10 mag safe
limit for saturation risk are considered.  These stars are further filtered
out so that their RAVE DR4 spectra (1) have S/N$\geq$45, (2) provided valid
atmospheric parameters ($T_{\rm eff}$, $\log g$, [M/H]) via both {\em sparv}
and {\em kordo} pipelines, (3) passed the chemical analysis pipeline of
\citet{boe11}, and (4) do not show peculiarities according to
\citet{mat12} pipeline.  About 51\,000 stars are plotted in
Figure~15.

For galactic latitudes $|b| \geq 25^\circ$, Figure~15 shows that there is a
negligible bulk difference between photometric temperatures (assuming the
slab reddening described above) and spectroscopic $T_{\rm eff}^{\rm sparv}$
temperatures for both red-giant/red-clump (RC) and main-sequence/turn-off
(MS/TO) stars, with no appreciable difference between \citet{zwi10} and
\citet{bin14} distances.  Most of the RAVE stars are out of the dust
slab, so their accurate distance does not matter here.  The rms of the
differences between photometric and spectroscopic temperature ($\Delta T$)
is 118~K for RC stars and 160~K for MS/TO stars, which are comparable to the
rms of spectroscopic temperatures derived independently from multiple RAVE
spectra of the same star.  The comparison between photometric and $T_{\rm
eff}^{\rm kordo}$ spectroscopic temperatures show a minimal bulk difference
for RC stars and instead a substantial one for MS/TO stars, while in both
cases the rms is similar to what found for $T_{\rm eff}^{\rm sparv}$.  The
effect of fixing $\log g$ and [M/H] to the spectroscopic values are explored
in the fifth row of Figure~15 (panels $e$, $e'$).  While somewhat reducing
the rms for both RC and MS/TO stars, it introduces an offset of about 40 K
for RC stars.

At lower Galactic latitudes ($|b|$$<$25$^\circ$), the simple slab reddening
shows its limitation on RC stars by introducing some bulk difference
between photometric and spectroscopic temperatures for both $T_{\rm
eff}^{\rm sparv}$ and $T_{\rm eff}^{\rm kordo}$, in addition to 
expanding the rms.  At these low Galactic latitudes, compensation for
reddening requires a more sophisticated approach, a true 3D model of the
reddening distribution that will be derived elsewhere in this series.

The effect of assuming null reddening for RAVE stars is explored in row
$f$,$f'$ of Figure~15.  The net effect is to lower the temperature of both
MS/TO and RC stars, because the redder energy distribution caused by
extinction requires cooler stellar models to be fitted.

The last row of Figure~15 (panels $g$,$g'$), compares spectroscopic and
photometric temperatures when all parameters are simultaneously fitted by
the photometric data and no assumption on the reddening is done.  The
comparison is good for RC stars, with no significant offset (at least
against $T_{\rm eff}^{\rm sparv}$) and an rms similar to the cases
when the reddening is fixed (first four rows of Figure 15).  For MS/TO stars
the $T_{\rm eff}$ so derived is instead unreliable, being characterized by a
large offset and rms.  The different behaviour between RC and MS/TO stars
can be traced to the fact that the maximum of the energy distribution for RC
stars falls well within the wavelength range covered by the
APASS+2MASS photometric bands, while for hotter MS/TO stars this maximum
tends to move to wavelengths shorter than covered by APASS+2MASS.

By adopting the atmospheric parameters from spectroscopy, the fit to
$B V g' r' i' J H K$ returns the reddening.  For RAVE DR4 stars this is
explored in Figure~16.

At higher galactic latitudes {\em sparv} RAVE DR4 atmospheric parameters
return a reddening which is very well in line with the {\em slab} reddening,
as illustrated in panel $b$ of Figure~16. No trend is present and the offset
is essentially null. Also {\em kordo} RAVE DR4 atmospheric parameters return
a null trend (panel $a$ of Figure~16), with a smaller rms but an appreciable
offset.  Closer to the galactic plane (panels $d$ and $e$ of Figure~16) the
comparison with the {\em slab} reddening shows the limitations of the {\em
slab} approach already commented upon above. The last row of Figure~16
compares (separately for low and high galactic latitudes) the reddening
derived by fixing the atmospheric parameters to the {\em sparv} values and
the reddening obtained when the atmospheric parameters are fitted
simultaneously with $E_{B-V}$. As noted above for $T_{\rm eff}$, the
comparison is excellent for high galactic latitude RC stars, with a null
trend and offset and only a marginal increase in rms. The comparison is less
favorable for MS/TO stars and for low galactic latitude RC stars.

Figure~17 displays the distribution of the differences between the
observed magnitudes and those of the best fitting synthetic stellar model,
for three representative panels of Figure~15 (for other panels they look
similar).  It is not tricking how the distributions for optical bands
are much sharper than those for infrared bands.  This reflects the fact that
optical bands carry more information (closer to the peak of the energy
distribution and where the effect of the reddening is larger) than infrared
bands, as well as the fact that the errors of 2MASS data are much
larger than those for APASS (cf.  Eq.1).

The much greater importance of optical bands compared to infrared ones in
constraining the temperature and reddening of stars is reaffirmed in
Figure~18. Here the same computation done for the first row of Figure~15
(its panels $a$,$a'$ are copied at the top of Figure~18 for comparison
purposes), are repeated by $\chi^2$ fitting only the APASS optical data or
only the 2MASS infrared data. It is quite obvious how in the case of
optical-only data the overall picture is retained, while for IR-only data
the fit is far less constrained.

\section{Conclusions}

The APASS all-sky, Landolt-Sloan $B$,$V$,$g'$,$r'$,$i'$ photometric survey
now provides the missing broad-band optical photometric support to the RAVE
spectroscopic survey, and coupled with 2MASS $J$,$H$,$K$ infrared data, it
offers a wide-wavelength coverage.  The high astrometric and photometric
accuracy of the APASS data has been verified on a large sample of
standard and literature stars.  With this first paper in the series, we
provide $B$,$V$,$g'$,$r'$,$i'$ data for the stars included in the latest
RAVE DR4 data release, and derive temperature and reddening for them under
different conditions by $\chi^2$ fitting to a densely populated synthetic
photometric library.  The comparison of photometric and spectroscopic
temperatures of RAVE stars shows a good agreement, so that photometric
temperature can be either valid {\em per se} or as {\em prior} to
spectroscopic analysis.  RAVE stars at Galactic latitudes $|b| \geq
25^\circ$ display an $E_{B-V}$ reddening quite well approximated by a
slab distribution of the interstellar dust that is characterized by 140
pc extensions on either side of the Galactic plane and a reddening at the
Galactic poles amounting to $E^{poles}_{B-V} =0.036$ mag.  Future papers in
this series will focus on RAVE peculiar stars, on the 3D structure of 
interstellar reddening as constrained by the APASS+2MASS photometry
of RAVE stars, and on the results of the APASS extention to brighter
magnitude limits and additional $u'$, $z'$ and $Y$ photometric bands.
 
\begin{acknowledgements}
The referee M. Bessell is acknowledged for the careful reading and relevant
suggestions.  Funding for APASS has been provided by the Robert Martin Ayers
Sciences Fund.  We also particularly recognize the support from the many
volunteers and vendors.  The APASS web site is at
http://www.aavso.org/apass.  Funding for RAVE has been provided by: the
Australian Astronomical Observatory; the Leibniz-Institut fuer Astrophysik
Potsdam (AIP); the Australian National University; the Australian Research
Council; the French National Research Agency; the German Research Foundation
(SPP 1177 and SFB 881); the European Research Council (ERC-StG 240271
Galactica); the Istituto Nazionale di Astrofisica at Padova; The Johns
Hopkins University; the National Science Foundation of the USA
(AST-0908326); the W.  M.  Keck foundation; the Macquarie University; the
Netherlands Research School for Astronomy; the Natural Sciences and
Engineering Research Council of Canada; the Slovenian Research Agency; the
Swiss National Science Foundation; the Science \& Technology Facilities
Council of the UK; Opticon; Strasbourg Observatory; and the Universities of
Groningen, Heidelberg and Sydney.  The RAVE web site is at
http://www.rave-survey.org
\end{acknowledgements}

\end{document}